\documentclass[article]{jss}

\usepackage{orcidlink,thumbpdf,lmodern}

\usepackage{framed}

\author{Irena Chen~\orcidlink{0000-0002-9366-8506}\\University of Michigan, Ann Arbor
\And Qiyuan Shi\\University of Michigan, Ann Arbor
\AND Scott L. Zeger~\orcidlink{0000-0001-8907-1603}\\The Johns Hopkins University, Baltimore
\And Zhenke Wu~\orcidlink{0000-0001-7582-669X}\\University of Michigan, Ann Arbor}
   
\Plainauthor{Chen I, Shi Q, Zeger SL, Wu Z}

\title{\pkg{baker}: An \proglang{R} package for Nested Partially-Latent Class Models}
\Plaintitle{\pkg{baker}: Nested Partially-Latent Class Models} 
\Shorttitle{\pkg{baker}: Nested Partially-Latent Class Models}

\Abstract{
This paper describes and illustrates the functionality of the \pkg{baker} {\sf R} package. The package estimates a suite of nested partially-latent class models (NPLCM) for multivariate binary responses that are observed under a case-control design. The \pkg{baker} package allows researchers to flexibly estimate population-level class prevalences and posterior probabilities of class membership for individual cases. Estimation is accomplished by calling a cross-platform automatic Bayesian inference software {\sf JAGS} through a wrapper \proglang{R} function that parses model specifications and data inputs. The \pkg{baker} package provides many useful features, including data ingestion, exploratory data analyses, model diagnostics, extensive plotting and visualization options, catalyzing communications between practitioners and domain scientists. Package features and workflows are illustrated using simulated and real data sets.
}

\Plainkeywords{keywords, not capitalized, Java}

\Address{
  Irena Chen, Qiyuan Shi, Zhenke Wu$^*$\\
  Department of Biostatistics\\
  University of Michigan\\
  1415 Washington Heights\\
  Ann Arbor, Michigan 48109, U.S.A.\\
  $^*$Corresponding author E-mail: \href{mailto:zhenkewu@umich.edu}{\nolinkurl{zhenkewu@umich.edu}}\\
  $^*$Corresponding author URL: \url{zhenkewu.com}\\
  
  Scott L. Zeger\\
  Department of Biostatistics\\
  The Johns Hopkins University\\
  615 N. Wolfe Street\\
  Baltimore, Maryland 21205, U.S.A.\\~\\
}

\usepackage{zwcommands}

\begin{document}
\section{Introduction}
\label{sec::intro}

This paper introduces the \pkg{baker} \proglang{R} package that estimates a suite of nested partially-latent class models (NPLCM) for multivariate binary responses that are observed under a case-control design. There are five popular \proglang{R} packages that provide functionalities to perform latent class analysis and some extensions on Comprehensive \proglang{R} Archive Network (CRAN) Task View of ``Cluster Analysis and Finite Mixture Models'' \citep{ctvcluster}. Our software is unique in its contribution to provide models and associated diagnostic and plotting functions for conducting Bayesian latent class analyses using data collected under a case-control design, where the primary statistical goal is to estimate the population- and individual-level class mixing weights among the cases. In particular, functions in \pkg{baker} implement recent methodological developments in \citet{wu2016partially}, \citet{wu2017nested}, and \citet{wu2021regression}. In practice, the package provides a simple interface that will allow researchers to reap the benefits of the NPLCMs via Markov chain Monte Carlo (MCMC) sampling without having to code the algorithms.

First formulated by \citet{lazarsfeld1950thelogical}, latent class models (LCMs) have become an important tool for modeling multivariate discrete responses \citep[e.g.,][]{Goodman1974, dunson2009nonparametric} and model-based clustering \citep[e.g.,][]{vermunt2002latent}. LCMs and various extensions have been used as primary workhorses driving discoveries and improved predictions in numerous scientific fields including psychology \citep[e.g.,][]{xu2017identifiability}, sociology \citep[e.g.,][]{mccormick2016probabilistic}, and public health \citep[e.g.,][]{stephenson2019robust}. There are currently several popular \proglang{R} packages that can perform general-purpose latent class analysis. The \pkg{poLCA} package developed by \cite{poLCA2011} provides a rich collection of functions to conduct latent class analysis for polytomous response variables and allows for the inclusion of regression variables to influence the class membership probabilities for each individual. Missing data is also handled under the assumption of missing at random. The \pkg{BayesLCA} package \citep{BayesLCA2014} provides functions for latent class analysis of multivariate binary responses in a Bayesian framework via expectation-maximization, MCMC, or variational Bayes. However, missing data is not handled in its current version (Version 1.9). In addition, we note that the models fitted by both  \pkg{poLCA} and \pkg{BayesLCA} make a classical local independence (LI) assumption for the multiple responses given class membership, which may be violated in many real-world settings. The \pkg{randomLCA} package provides functions to fit LCMs with optional random effects that cause local dependence (LD), of which LI is a special case.

The \pkg{baker} package provides multiple novel advantages to existing software. First, \pkg{baker} enables case-control analyses with or without covariates in the NPLCM framework. The case-control design is vital to valid scientific inference in many large-scale clinical and biomedical applications. For example, in the largest pediatric pneumonia etiology study to date \citep{o2019aetiology}, biological samples are collected from subjects with clinically-defined disease (``cases'') and subjects without disease (``controls''). Panel molecular diagnostic tests targeting multiple putative disease-causing agents may be performed on the collected samples, resulting in multivariate binary data under a case-control design. The control subjects have the observed class of not having the said disease. Their data serve as statistical control to estimate the measurement specificity when interpreting the error-prone test results in the cases. 

Second, \pkg{baker} can fit models under deviations from the classical LI assumption in latent class analyses. Different from the continuous random effects approach taken in \citet{Qu1996}, the methodology implemented by \pkg{baker} uses a parallel factor decomposition with a stick-breaking prior to enable parsimonious approximation of potential LD between the multivariate binary responses given class membership. This enables faster computation and data-driven learning of empirical LD structures. 

Third, \pkg{baker} is designed to handle multiple sets of case-control or case-only measurements of distinct degrees of measurement error. These measurements are classified into two broad types: 1) bronze-standard (BrS) data that are available for both cases and controls but with imperfect sensitivity or specificity; and 2)  silver-standard (SS) data that are only available among cases, with perfect specificity but imperfect sensitivity. The \pkg{baker} can also work under missing data under the assumption of missing at random.

Finally, the \pkg{baker} package conducts full Bayesian inference via MCMC by calling a cross-platform and versatile automatic Bayesian software \proglang{JAGS} \citep{plummer2003jags,JAGSsourceforge} via a wrapper function \code{baker::nplcm()} that parses model specifications and data inputs (see Section \ref{sec:software} for the software design choice). The main quantities of interest are 1) the population-level class prevalences and 2) posterior probabilities of class membership for the individual cases. The \textbf{baker} package quantifies the uncertainty associated with these estimates and provides numerical and graphical summaries to assist in interpreting and communicating these results.

The remainder of this paper is organized as follows. Section \ref{sec:model} provides a brief overview of the NPLCM framework as a case-control extension of the classical LCMs. The NPLCM likelihood and prior specifications are detailed in Section \ref{sec::nplcm} without explanatory covariates; Section \ref{sec::nplcm_reg} covers the regression extension. Section \ref{sec::mcmc} discusses model fitting and diagnostics. Section \ref{sec:software} gives a brief description of workflows and code underlying \pkg{baker}. This is followed in Section \ref{sec::illustration} by analyses of simulated and real data sets that demonstrate many of the package's features. Finally, Section \ref{sec:conclusions} summarizes the main advantages of the \pkg{baker} package relative to existing software and future developments to expand its utility.

All figures in this paper can be reproduced by following the user vignette provided along with this article. The stable version of the package is available via CRAN (\url{https://CRAN.R-project.org/package=baker}); the development version can be accessed at \url{https://github.com/zhenkewu/baker}. 


\section{Model}
\label{sec:model}
\subsection{Latent class models: A brief review}
\label{sec:term}

Latent class models (LCMs) for discrete latent and discrete manifest variables were developed and widely applied since the 1950s \citep[e.g.][]{lazarsfeld1950thelogical, Goodman1974}. LCMs constitute a family of distributions for correlated discrete measurements. The conventional LCM generally makes \textit{local independence} (LI) assumption that manifest variables are independent of one another given the latent class. In the multivariate binary case, individual $i$'s measurement vector, $\bm{M}_i  = (M_{i1}, ..., M_{iJ})^\top$, is linked to her latent class ($I_i$) by the simple product likelihood $\PP(\bM_{i} \mid I_i  = \ell, \bm{\theta})= \prod_{j=1}^J \PP(M_{ij} \mid I_i = \ell, \bm{\theta})$, where $I_i$ takes value from $\{1,\ldots, L\}$ and  ${\bm{\theta}}$ represents the collection of measurement parameters --- sensitivities and specificities. We then obtain the observed likelihood by summing over all the possible values of $I_i$, i.e., $\PP(\bM_i  \mid \bm{\theta}, \bm{\pi}) = \sum_{\ell=1}^L \pi_{\ell} \prod_{j=1}^J \PP(M_{ij} \mid I_i=\ell, \bm{\theta})$, where ${\bpi}$ is a vector of mixing weights of length $L$. The LI assumption implies that the latent membership $I_i$ completely explains the marginal dependence in $\bm{M}_i$. Under local identifiability conditions \citep{allman2009identifiability}, we can estimate $\bpi$ and $\btheta$ by the values that optimally reduce the observed dependence among measurements given latent class, e.g., through the expectation-maximization (EM) algorithms. Individual classification can then proceed by applying Bayes rule using the estimated parameters.

Below, we introduce the NPLCM family of models using case-control BrS measurements obtained from a single source (referred to as a ``slice'' in the \pkg{baker} package). In Section \ref{sec:multiple_slices_BrS_SS}, we generalize the model to using multiple slices of BrS measurements, and to integrating case-only SS measurements which are special cases of BrS measurements having false positive rate of zero (perfect specificity).

\subsection{Data structure and notation for case-control studies}
\label{sec::data_notation}
Let $Y_i=1$ indicate a case subject with the clinically-defined disease and  $Y_i=0$ indicate a control subject without disease. Let $\bm{M}_{i} = (M_{i1},...,M_{iJ})^\top\in\{0,1\}^J$ represent the multivariate binary case-control, non-gold-standard diagnostic test results from subject $i$. Let $\cD = \{(\bM_i, Y_i, \bX_iY_i, \bW_i), i=1, \ldots, N\}$ represent data, where $\bX_i=(X_{i1}, \ldots, X_{ip})^\top$ are the $p$ primary covariates in CSCF functions and hence must be available for cases, and $\bW_i = (W_{i1}, \ldots, W_{iq})^\top$ are $q$ covariates that are available in the cases and the controls. $\bX_i$ and $\bW_i$ may be identical, overlapping or completely different. $\bX_iY_i=\bX_i$ for a case $Y_i=1$; $\bX_iY_i$ is a vector of zeros for a control subject. For notational convenience, we have ordered the continuous variables, if any, in $\bX_i$ and $\bW_i$ as the first $p_1$ and $q_1$ elements, respectively. In this paper, we focus on pre-specified $\bX_i$ and $\bW_i$.

\subsubsection{Classes defined by latent states}
\label{sec:latent_state_notation}
We first introduce notation for the true but unobserved latent classes (e.g, causes of disease) among the case subjects. Suppose a total of $J$ ``agents'' or ``items'' are measured by the diagnostic tests. Let a binary variable $\iota_{ij}$ indicate whether ($\iota_{ij}=1$) or not ($\iota_{ij}=0$) the $j$-th agent caused case $i$'s disease. We also allow more than one agent to cause the disease. We therefore have $\bm{\iota}_i = (\iota_{i1}, \ldots, \iota_{iJ})^\top\in \{0,1\}^J$ which is a vector of multiple binary indicators that represents the causes for subject $i$. We will also refer to $\bm{\iota}_i$ as ``latent states'' for case subject $i$. Note that we allow the all-zero latent states $\bm{\iota}_i = \bm{0}_{J\times 1}$ to represent a case with a ``Not Specified'' (NoS) cause. For example, in the Pneumonia Etiology Research for Child Health (PERCH) study, ``NoS'' can represent the subgroup of cases whose diseases are caused by agents not specified as molecular targets in the diagnostic tests (such as polymerase chain reaction, PCR). We will refer to cases having the same pattern of multivariate binary pattern $\bm{\iota}_i$ as belonging to the same ``disease class'' or ``class'' for short.

In this paper, we assume that there are $L$ classes of \textit{pre-specified} latent state patterns (possibly ``NoS'') among the cases. Let the set $\cA$ comprise the \textit{pre-specified} distinct multivariate binary patterns so that $|\cA| = L$, where $|\cA|$ is the cardinality of $\cA$. We then introduce class indicators by arbitrarily labeling elements in $\cA$ from $1$ to $L$. We can now use $I_i$ that takes value from $\{1,\ldots, L\}$ to indicate case subject $i$'s class. We also let $\cC_\ell=\{j: \iota_{ij}=1, I_i = \ell, j=1, \ldots, J\}$ represent the subset of causative agents for disease class $\ell$; for the NoS class, we have $\cC_{\sf NoS}= \emptyset$.

For a control $i'$, we use $I_{i'}=0$ to indicate $\bm{\iota}_{i'}=\zero_{J\times 1}$, e.g., no lung infection in the PERCH study. For a case or a control, the value of $I_i$ thus corresponds to a particular state pattern, so we can write $\bm{\iota}_i = \bm{\iota}_{i}(I_i)$.

To illustrate the scientific meaning of the notation, consider a hypothetical list of $J=5$ species of pathogens (``items'') in the context of PERCH study; they are targeted by the panel diagnostic tests. First, under the assumption that there are only single-pathogen causes and no NoS class, we have $L=J=5$ disease classes with distinct patterns of $\bm{\iota}$: 
\[\cA = \{(1, 0,0,0,0)^\top, (0, 1,0,0,0)^\top, \cdots, (0, 0,0,0,1)^\top\}.\] We can label the five disease classes by $1, \ldots, L=5$, so that, for example, $I_i=2$ corresponds to $\bm{\iota}_i=(0,1,0,0,0)^\top$ and $\cC_2 = \{2\}$, $I_i=5$ corresponds to $\bm{\iota}_i=(0,0,0,0,1)^\top$ and $\cC_5 = \{5\}$. Second, under a less restrictive assumption of single- or double-pathogen causes (still no NoS class), we have $L={J\choose1}+{J\choose2}=5+10=15$ disease classes. For example, cases with the first and the third pathogen infecting the lung are represented by $\bm{\iota}_{i}=(1,0,1,0,0)^\top$. It has the subset of causative agents $\cC_\ell=\{1,3\}$ where $I_i = \ell$ is an arbitrary integer label of the disease class.

\subsection{Nested partially latent class models for case-control studies}
\label{sec::nplcm}


\citet{wu2016partially} and \citet{wu2017nested} introduce a generalization to the latent class model in order to address two aspects of our particular setting. First, the latent classes are called ``partially latent'' since class membership is known for the subset of controls, but not cases. Second, the conditional independence assumption is relaxed by allowing for \textit{nested subclasses} within each class. The inclusion of subclasses accounts for the possibility of correlation or dependence between measurements. The \textbf{baker} package implements this \textit{nested partially latent class model} (NPLCM) framework. For ease of interepretation, we present the models by referring to terminologies in the PERCH study.

\subsubsection{Likelihood}

The NPLCM model likelihood can be specified via the following generative processes for the controls and the cases, respectively.
 \begin{alignat}{3}
 & {\sf control~subclass:}   \quad && Z_i \mid Y_i=0 \sim  {\sf Categorical}_K\left\{ \bnu\right\}, \bnu  \in \cS_{K-1}, \label{eq:control_nplcm1} \\
                                                & {\sf control~data:}   \quad && M_{ij}\mid \iota_{ij}=0, Z_i = k\sim  {\sf Bern}\left\{\psi^{(j)}_k\right\},  \text{~independently~for~}j=1,...,J, & \label{eq:control_nplcm2}
                \end{alignat}
\noindent where $\bnu=(\nu_1, \ldots, \nu_K)^\top$ is the vector of subclass probabilities and lies in a probability simplex. When $K=1$, the model is referred to as PLCM \citep{wu2016partially}.  Let $\bPsi = \{\psi_k^{(j)}\in (0,1)\}$ be a $J\times K$ matrix comprising false positive rates (FPRs), which are necessary for modeling the imperfect binary measurements among the controls. Let $\bpsi^{(j)}$ and $\bpsi_k$ represent the $j$-th row and $k$-th column. The data generating process for cases is as follows, with an additional Step (\ref{eq:case_subclass1}) for drawing  a subclass indicator $Z_i$ for each case subject:
 \begin{alignat}{3}
                                 & {\sf disease~class:}   \quad && I_i \mid  Y_i=1\sim  {\sf Categorical}_L\left\{\bpi\right\}, \bpi \in \cS_{L-1},\label{eq:case_class1}\\
                                 & {\sf case~subclass:}   \quad && Z_i \mid Y_i=1 \sim  {\sf Categorical}_K\left\{\bEta\right\}, \bEta\in \cS_{K-1},  \label{eq:case_subclass1}\\
                                                       &   {\sf convert~class~to~states:}  \quad && \bm{\iota}_{i} = \bm{\iota}_{i}(I_i) \in \cA; \nonumber \\   
                & {\sf case~data:}   \quad && M_{ij}\mid \iota_{ij}, Z_i = k, I_i = \ell, \sim  {\sf Bern}\left\{p_{k\ell}^{(j)}\right\}, \text{~independently~for~}j=1,...,J, & \label{eq:cause_meas}\\
                & {\sf response~probabilities} : \quad &&p_{k\ell}^{(j)}=\begin{cases}
    \theta_k^{(j)}, & \iota_{ij}=1;\\
    \psi_k^{(j)}, & \iota_{ij}=0,
  \end{cases} ~\quad k=1, \ldots, K, \text{~and~}\ell = 1, \ldots, L.\label{eq:case_positive_rate_nplcm}
                \end{alignat}
At Step (\ref{eq:case_subclass1}), the NPLCM introduces $K$ unobserved subclasses with weights $\bEta = (\eta_1, \ldots, \eta_K)^\top$. The weights are shared across $L$ disease classes. Let $\bTheta = \{\theta_k^{(j)}\in(0,1)\}$ be a $J\times K$ matrix where $\theta^{(j)}_k$ represents the positive response probability in subclass $k$ if item $j$ is causative in a disease class. We also refer to $\theta^{(j)}_k$ as true positive rate (TPR) or sensitivity as in 
PLCM. Let $\btheta^{(j)}$ and $\btheta_k$ represent the $j$-th row and $k$-th column. In Step (\ref{eq:case_positive_rate_nplcm}),  $p_{k\ell}^{(j)}$ represents the positive response probability of $M_{ij}$ in subclass $k$ of disease class $\ell$, which equals the TPR $\theta_k^{(j)}$ for a causative pathogen and the FPR $\psi_k^{(j)}$ otherwise. We collect all the positive response probabilities for subclass $k$ in disease class $\ell$  into $\bp_{k\ell} = (p_{k\ell}^{(1)}, \ldots, p_{k\ell}^{(J)})^\top$.

The population-level class prevalences in the cases are referred to as ``cause-specific case fractions''(CSCF): $\boldsymbol{\pi}= (\pi_{1}, \ldots, \pi_{L})^\top$, which represent the population-level distribution of disease classes. $\bpi$ is the population-level class prevalences among the cases and is often of primary scientific interest. $\bpi$ is also referred to as cause-specific case fractions \citep[CSCFs][]{wu2021regression}.

\subsubsection{Prior}
For NPLCM, we specify the prior distributions on unknown parameters as follows:
\begin{eqnarray}
\bm{\pi} & \sim & {\sf Dirichlet}(a_1,\dots,a_{L}),\\
\psi_k^{(j)} &\sim& {\sf Beta}(b_{1kj},b_{2kj}), j = 1, ..., J; k\leq K,\\
\theta_k^{(j)} & \sim & {\sf Beta}(c_{1kj},c_{2kj}), j=1,...,J; k\leq K,\\
\eta_k &\sim&  U_k\prod_{s<k}\left[1-U_{s}\right], ~~~~U_k\sim {\sf Beta}(1,\alpha_1), k<K; U_K=1;\label{eq:case_sb}\\%
\nu_k &\sim & V_k\prod_{s<k}[1-V_{s}], ~~~~V_k\sim {\sf Beta}(1,\alpha_0), k<K; V_K=1; \label{eq:ctrl_sb}\\
 \alpha_0, \alpha_1 &\sim&{\sf Gamma}(0.25,0.25),\label{eq:stick.break.hyperprior}
\end{eqnarray}
where prior independence is also assumed among these parameters. As discussed in more detail by \citet{wu2016partially}, the NPLCM likelihood similarly has the TPRs $\bTheta$ that are not fully identified by the model likelihood and hence is partially identified \citep{Jones2010}. Therefore, we choose $(c_{1kj},c_{2kj}), \forall k,j$, so that the $2.5\%$ and $97.5\%$ quantiles of the Beta distribution with parameters $(c_{1kj},c_{2kj})$ match the prior minimum and maximum TPR values elicited from domain experts. Otherwise, we use the default value of $1$s for the Beta hyperparameters. Hyperparameters for the etiology prior, $(a_1,...,a_J)^\transp$, are usually $1$s to denote equal and flat prior weights for each disease class if expert prior knowledge is unavailable. Finally, in (\ref{eq:case_sb}) and (\ref{eq:ctrl_sb}), we have specified truncated stick-breaking priors for both $\bm{\eta}$ and $\bnu$ that on average place decreasing weights on the $k$th subclass as $k$ increases \citep{sethuraman1994constructive}.

\subsection{Regression extensions of NPLCM}
\label{sec::nplcm_reg}

\subsubsection{Likelihood}
An extension of the NPLCM allows for covariates to predict latent class membership by allowing the priors of the CSCFs and the subclass mixing weights to be a function of the observed explanatory variables.  There may be biological or epidemiological support to include covariates in the likelihood function. For example, date of diagnosis may be informative if the disease of interest is known to have seasonal patterns. 
\par We let the CSCFs depend on $\bX_i$ by using a classical multinomial logistic regression:
\begin{align}
\pi_{i\ell}  & = \pi_\ell(\bX_i)= \exp\{\phi_\ell(\bX_i)\}/\sum_{\ell'=1}^L\exp\{\phi_{\ell'}(\bX_i)\}, \ell =1, ..., L, \label{eq:cscf_reg}
\end{align}
where  $\phi_{\ell}(\bX_i)-\phi_L(\bX_i)$ is the log odds of case $i$ in disease class $\ell$ relative to $L$: $\log {\pi_{i\ell}}/{\pi_{iL}}$. We treat all the disease classes symmetrically in this formulation, which simplifies the prior specification. 

The regression extension assumes the control subclass weights are covariate-dependent:
\begin{alignat}{3}
                & {\sf Extend~ (\ref{eq:control_nplcm1}) - control~subclass:}   \quad && Z_i \mid \bm{W}_i, Y_i=0 \sim  {\sf Categorical}_K\left\{ \bnu_i\right\}, \bnu_i = \bnu(\bW_i) \in \cS_{K-1}, \label{eq:control_subclass1}
                \end{alignat}
where, as in NPLCM \citep{wu2017nested}, the subclass indicators $Z_i$'s are nuisance quantities for inducing dependence among the multivariate binary responses $\bM_i$, but now given covariates.  $\bnu_{i} =\left(\nu_{i1}, \ldots, \nu_{iK}\right)^\top$ is the vector of control subclass probabilities that now may depend on $\bW_i$. Scientifically, we are not interested in how the subclass probabilities are associated with covariates. We introduce $\bnu(\bW)$ here because, upon integrating over the distribution of $Z_i$ in (\ref{eq:control_subclass1}), it helps define a flexible conditional distribution of $\bM_i$ given covariates $\bW_i$. 

For cases, we follow the case model for NPLCM, but extend in two aspects:  let CSCFs depend on covariates $\bX_i$ and let case subclass weight depend on covariates $\bW_i$. That is,
                \begin{alignat}{3}
                                 & {\sf Extend~ (\ref{eq:case_class1}) -disease~class:}   \quad && I_i \mid \bm{X}_i , Y_i=1\sim  {\sf Categorical}_L\left\{\bpi_i\right\}, \bpi_i=\bpi(\bm{X}_i) \in \cS_{L-1},\label{eq:case_class}\\
                & {\sf Extend ~(\ref{eq:case_subclass1}) -case~subclass:}   \quad && Z_i \mid \bm{W}_i, Y_i=1 \sim  {\sf Categorical}_K\left\{\bEta_{i}\right\}, \bEta_i = \bEta (\bW_i)\in \cS_{K-1},  \label{eq:case_subclass}
                \end{alignat}
where $\bpi(\bX_i)= (\pi_1(\bX_i), \ldots, \pi_L(\bX_i))^\top$ are CSCF functions evaluated at $\bX_i$, and $\bEta(\bW_i) =\left(\eta_{i1}(\bW_i), \ldots, \eta_{iK}(\bW_i)\right)^\top$ is the vector of case subclass probabilities evaluated at $\bW_i$. Both $\bpi_i$ and $\bEta_i$ are quantities from probability simplexes. 

\subsubsection{Detailed regression specification}

\noindent \underline{\textit{CSCF regression}.}
We further assume additivity in a partially linear model:
\begin{align}
\phi_\ell(\bm{x}; \Gamma_\ell^\pi) = \sum_{j=1}^{p_1} f^\pi_{\ell j}(x_j; \bbeta^\pi_{\ell j})+\tilde{\bx}^\top\bgamma^\pi_\ell,
\end{align}
where $\tilde{\bx}$ is the subvector of the predictors $\bx$ that enters the model for all disease classes as linear predictors which may include an intercept, and $\Gamma_\ell^\pi = [(\bbeta^\pi_{\ell 1})^\top,\ldots, (\bbeta^\pi_{\ell p_1})^\top, (\bgamma_\ell^\pi)^\top]^\top$ is the vector of the regression coefficients for disease class $\ell$. For covariates such as enrollment date that serve as proxy for factors driven by seasonality, non-linear functional dependence is expected. We approximate unknown functions of a standardized continuous variable such as $f^{\pi}_{\ell j}$ via basis expansions and along with  a prior on the basis coefficients to encourage smoothness.

\noindent \underline{\textit{Control subclass weight regression}.} We specify $\nu_{ik}$ by logistic stick-breaking parameterization:
\begin{align} 
\nu_{ik} & = g(\alpha^\nu_{ik})\prod_{s<k} \{1-g(\alpha^\nu_{is})\}, \text{~if~}k<K, \text{~and~} \prod_{s<k} \{1-g(\alpha^\nu_{is})\} \text{~otherwise}, \text{~where~} \label{eq:lsbp}\\
\alpha^\nu_{ik}  & = \alpha^\nu_{k}(\bW_i = \bm{w};\Gamma^\nu_k) = \mu_{k0}+\sum_{j=1}^{q_1}f^\nu_{kj}(w_j; \bbeta^\nu_{kj})+\tilde{\bw}^\top\bgamma^\nu_k, \textrm{~for~} k=1, \ldots, K-1.\label{eq:lsbp_linpred}
\end{align} 
Let $\Gamma^\nu_k = [(\bbeta^\nu_{k1})^\top, \ldots, (\bbeta^\nu_{kq_1})^\top, (\bgamma^\nu_{k})^\top]^\top$ be the regression coefficients in the $k$-th subclass, and $\alpha^\nu_{ik}$ is subject $i$'s linear predictor at stick-breaking step $k=1, \ldots, K-1$; $g(\cdot): \RR \mapsto [0,1]$ is a link function. In the \pkg{baker} package, we use the logistic function $g(\alpha) = 1/\left\{1+\exp(-\alpha)\right\}$ which is consistent with (\ref{eq:cscf_reg}) so that the priors of the coefficients $\Gamma_k^\nu$ and $\Gamma_\ell^\pi$ can be similar. 





\noindent \underline{\textit{Case subclass weight regression}.} The case subclass weight curve $\bEta_k(\bW)$ is also specified via a logistic stick-breaking regression as in the controls but with different linear predictors $\alpha^\eta_{ik}$: $\eta_{ik} = g(\alpha^\eta_{ik}) \prod_{s<k} \{1-g(\alpha_{is}^\eta)\}$, $\forall k=1, \ldots, K-1$; $\eta_{iK}=\prod_{s<K} \{1-g(\alpha^\eta_{is})\}$. Given $\bTheta$ and $\bPsi$, $\bEta_k(\bW)$ fully determines the joint distribution $[\bM\mid \bW, I=\ell\neq0, \bTheta, \bPsi]$. We do not assume  $\eta_k(\bw)=\nu_k(\bw), \forall \bw$. Consequently, relative to the controls, the individuals in disease class $\ell$ may have different strength and direction of observed dependence between the causative $\{M_{j}: j\in \cC_\ell\}$ and non-causative $\{M_{j}: j\notin \cC_\ell\}$ pathogens, or between the non-causative pathogens. Let the $k$-th linear predictor
\begin{align} 
\alpha_{ik}^\eta & = \alpha_{k}^\eta(\bW_i=\bw; \Gamma_k^\eta) =  \mu_{k0}+\sum_{j=1}^{q_1} f^\eta_{kj}(w_j; \bbeta_{kj}^\eta)+\tilde{\bw}^\top\bgamma_k^\eta,
\end{align}
where $f_{kj}^\eta$ and $f_{kj}^\nu$ (from the control model) share the basis functions but the regression coefficients $\Gamma^\eta_k = [(\bbeta_{k1}^\eta)^\top,\ldots, (\bbeta_{kq_1}^\eta)^\top, (\bgamma_k^\eta)^\top]^\top$ differ from the control counterpart ($\Gamma_k^\nu$). In addition, we have used the same intercepts $\{\mu_{k0}\}$ in (\ref{eq:lsbp_linpred})  to ensure only important subclasses in the controls are used in the cases. For example, absent covariates $\bW$, a large and positive $\mu_{k0}$ effectively halts the stick-breaking procedure at step $k$ for the controls. This is because the $k$-th stick-breaking will take almost the entire remaining stick, resulting in $\nu_{k+1}$ that is approximately zero. Applying the same intercept $\mu_{k0}$ to the cases makes $\eta_{k+1}\approx 0$.

Section \ref{sec::illustration} provides examples of how to include discrete and continuous covariates in the model specification using the \textbf{baker} package.
 
\subsubsection{Priors}
\label{sec::priors}

The number of parameters in the model likelihood for the regression model \\$(\{\Gamma_\ell^\pi\}, \{\Gamma_k^\eta\}, \{\Gamma^\nu_k\},\{\mu_{k0}\}, \bTheta, \bPsi)$ is $\cO(LC_{\max}p_1+KC_{\max}q_1+JK)$ where $C_{\max}$ is the maximum number of basis functions in $\{f^{\pi}_{\ell j}, f_{kj}^\nu, f_{kj}^\eta\}$. It easily exceeds the number of observed distinct binary measurement patterns. To overcome potential overfitting and increase model interpretability, we \textit{a priori} encourage the following two features: ({a}) few non-trivial subclasses  uniformly over $\bW_i$ values, and ({b}) constant subclass weights over $\bW_i$ values $\eta_k(\cdot)=\eta_k$ and $\nu_k(\cdot)=\nu_k$. See \citet{wu2021regression} for the exact technical specifications.

\subsection{Posterior inference via MCMC}
\label{sec::mcmc}

We perform posterior inference via Markov chain Monte Carlo (MCMC) algorithm that draws posterior samples of the unknowns to approximate their joint posterior distribution \citep{Gelfand1990}. Flexible posterior inferences about any functions of the model parameters and individual latent variables are available by plugging in the posterior samples of the unknowns. All the models presented so far are available in the \pkg{baker} package. See \citet{wu2016partially,wu2017nested,wu2021regression} for details of the sampling algorithm.

\section{Software: design features and main function}
\label{sec:software}

\begin{figure}[htp]
    \centering
      \includegraphics[width=0.9\textwidth]{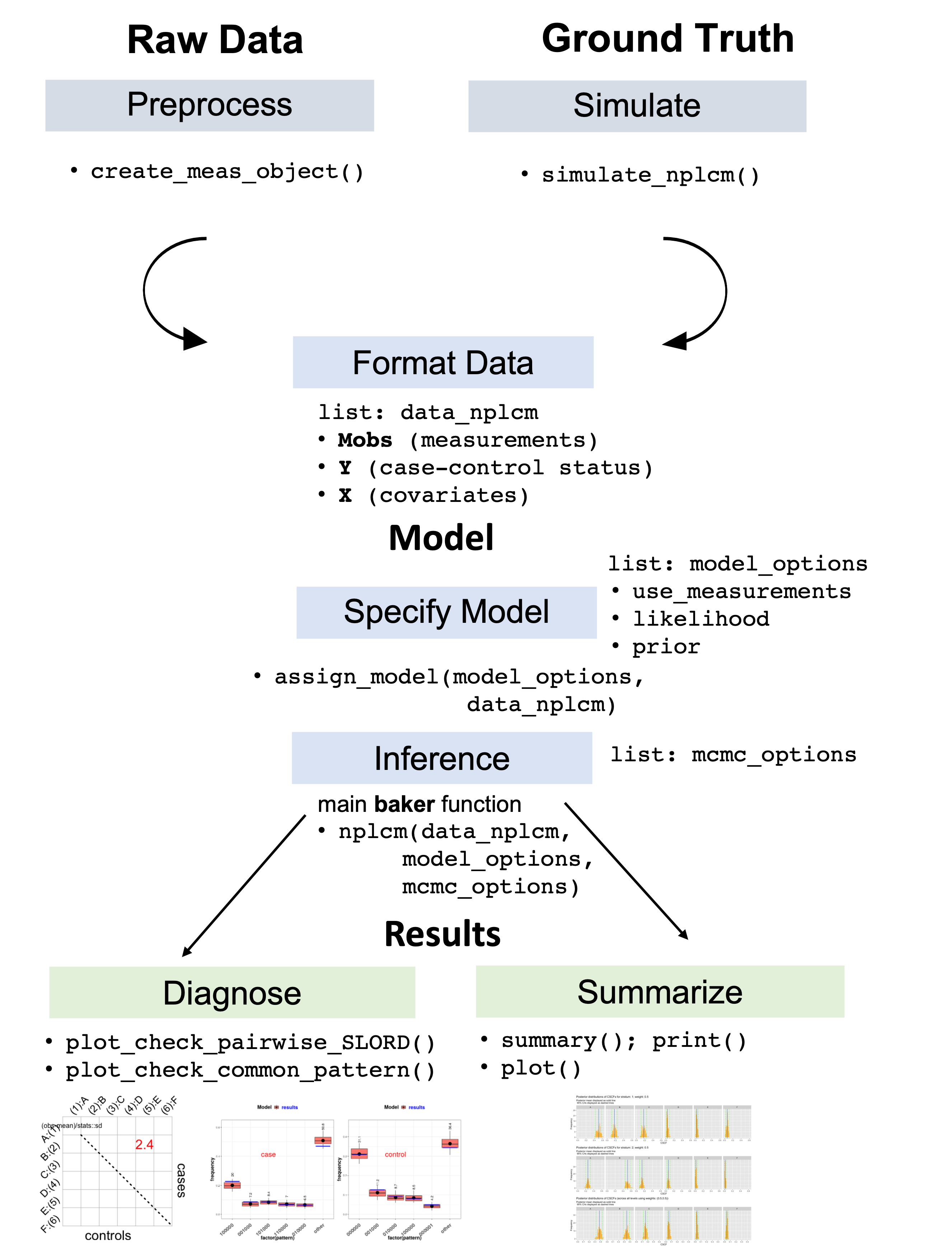}
    \caption{\pkg{baker} package workflow.}
    \label{fig:heuristic_baker}
\end{figure}

The Bayesian method for estimating population-level class prevalences and posterior probabilities of class memberships for individual cases is implemented by connecting \proglang{R} with another freely available cross-platform automatic Bayesian fitting program \citep[\proglang{JAGS} 4.2.0][]{JAGSsourceforge}. Figure \ref{fig:heuristic_baker} shows a schematic workflow that connects some \pkg{baker} functions and arguments to the steps in a data analysis or simulation pipeline. The \pkg{baker} package implements both exploratory and model-based analyses of case-control multivariate binary data. The package enables an analyst to organize multivariate binary diagnostic test results by their measurement standards (BrS or SS) and to calculate summaries such as the marginal positive rates for each item in the cases and controls; pairwise odds ratios can also be computed and visualized. The analyst can then specify which subsets of measurement data to use, the model likelihood, and the prior distributions for true positive rates and population-level CSCFs, among other model components. Based on these model specifications, \proglang{R} calls and instructs \proglang{JAGS} to fit the corresponding model to the data, performs model diagnostics, and stores the posterior results for ensuing inference of the key unknown quantities, such as the population-level class prevalences and class membership for individual cases. Finally, the package offers numerical and graphical summaries to display the evidence in the data and to facilitate model criticism.

The \pkg{baker} package uses the \proglang{JAGS} program to fit the specified NPLCMs; pre-installation of \proglang{JAGS} is required - the accompanying vignette contains detailed instructions about setting up \proglang{JAGS} for \pkg{baker} along with other required \proglang{R} package dependence.

\code{nplcm()} is the main function of the \pkg{baker} package and takes in three required arguments:
\begin{itemize}
\item \code{data_nplcm}: a named list containing data consisting of the following: 1) measurements \code{Mobs} - a named list containing \code{MBS} for BrS measurements and \code{MSS} for SS measurements, 2) case-control status \code{Y}, and 3) covariates \code{X};
\item \code{model_options}: a named list that specifies the data sources, model likelihood, and prior distributions for the model parameters;
\item \code{mcmc_options}: a named list that specifies how to set up the MCMC sampling algorithm for posterior inference.
\end{itemize}

We will provide detailed examples for each of the three arguments in the next section. The output of \code{nplcm()} is an object of class \code{"nplcm"} which contains the path to where results were stored (accessible via \code{\$DIR_NPLCM}) and the sampled values of model parameters which can be further manipulated via external posterior processing packages such as \pkg{coda} and \pkg{ggmcmc} (see Section \ref{sec:postprocess}). It is designed such that intermediate model results, model specifications, input data are retained for debugging and re-purposing for analyses not included in \pkg{baker}, such as post-stratification of CSCFs by discrete covarates (e.g., age group) using model results obtained from an NPLCM fitted without covariates. In addition, in high-performance computing, we may organize simulation settings by folder with proper names indicating the differences in the ground truth. Separate functions can be written and applied to these folders to obtain simulation results for various comparisons. Although the downside is the extra storage of results in the folder (and thus, the cost of additional disk memory),  we believe that retaining this information is often more beneficial in complex substantive applications. Fortunately, generic functions in the \pkg{baker} package can read and organize these information if the fitted object is provided.

\subsection{Multiple slices of BrS data and SS data}
\label{sec:multiple_slices_BrS_SS}

\code{nplcm()} can readily integrate more than one source of BrS measurements by supplying data to the argument \code{data_nplcm} and using data as specified in the argument \code{model_options}. For example, in the PERCH study, besides the NPPCR test for bacteria and viruses, pleural fluid PCR on the same set of pathogen targets may be performed; they are obtained from a different specimen with the same technology PCR and have different TPRs and FPRs.  In the argument \code{data_nplcm} (a list), we can add these measurements to the list \code{data_nplcm\$Mobs\$MBS} which itself may contain multiple elements; we refer to each source of BrS measurements as a ``slice''. The \pkg{baker} package can integrate multiple slices of BrS measurements. In addition, case-only SS measurements may be available, e.g., blood culture results on the subset of bacteria. SS measurements are assumed to have perfect specificity, i.e., measurements on controls are assumed to never return positive results; SS data are by definition case-only. Similarly, one may add SS data into the list \code{data_nplcm\$Mobs\$MSS} which itself may contain multiple elements or ``slices''. The model likelihood with additional BrS and/or SS data will be modified automatically by the \code{nplcm()} function. In the argument \code{model_options} (a list), we simply set the element \code{use_measurements = "BrS"} (\code{"SS"}) to use all slices of the provided BrS (SS) data; setting \code{use_measurements = c("BrS","SS")} will use both BrS and SS data for model estimation. We illustrate these data source specifications in Section \ref{sec:model_specification}.

\section{Illustrations}
\label{sec::illustration}

In this section, we give code snippets with explanations for data simulation, model specification, fitting, and numerical and graphical summaries of model results. By using simulated and a real data set, we illustrate the practical functionalities of the \pkg{baker} package. See the reproducible \code{RMarkdown} file provided along with this article for more illustrative examples of standard workflows.

Below, we first illustrate the three required arguments of the main function \code{nplcm()}: \code{data_nplcm} (Section \ref{sec:data_nplcm}), \code{model_options} (Section \ref{sec:model_specification}), \code{mcmc_options} (Section \ref{sec:mcmc_options}). Second, we fit specified models to illustrate the outputs of the main function \code{nplcm()}. Finally, we demonstrate how to use functions in the \pkg{baker} package to produce numerical and graphical summaries of the model results. Section \ref{sec:simulate_with_covariates} considers data simulation with covariates.

\subsection{Setting up data inputs: Simulation and structure}
\label{sec:data_nplcm}

In the following, we simulate and store BrS and SS measurement data to be used in the argument \code{data_nplcm}. \code{simulate_nplcm()} is a function that takes in the data generating parameters stored in a named list (e.g., \code{"set_parameter_noreg"} below) and outputs a data set containing 1) measurements (\code{Mobs}), which itself is a list with an element \code{MBS} that stores BrS case-control measurements and another element \code{MSS} that stores SS case-only measurements, and 2) case-control status (\code{Y}). 

Here, we provide an example of relevant parameters that are useful in simulating data. We illustrate the values that these model parameters can take. We need to specify the number of items in a slice of bronze-standard measurements (\code{J.BrS = 7}), the number of items in a slice of silver-standard measurements (\code{J.SS}), and the number of subclasses (\code{K}). In the code snippet below, we simulate data for \code{Nd = 300} cases and \code{Nu = 300} controls. \code{cause_list} specifies the names of the true causes, with \code{etiology} specifying the population-level proportions of cases due to each cause (CSCFs). The BrS data used to infer the classes are case-control measurements on six items \code{c("A","B","C","D","E","F")} which in this example happens to be the exact set of agents that can cause the disease; in general, the measured items may include non-causative items or miss causative items. SS data are case-only and can measure fewer items targeted by BrS measurements. For the BrS data, true (\code{ThetaBS}) and false positive rates (\code{PsiBS}) for all the subclasses must be specified; they are of identical dimensions
(\code{J.BrS} by \code{K}). In addition, one needs to specify two possibly different vectors of subclass weights for the case (\code{Eta}) and the control populations (\code{Lambda}), respectively; they determine the conditional dependence structure of the BrS measurements in the cases and the controls. For SS data, the false positive rates (\code{PsiSS}) must be all zeros indicating perfect specificity; the true positive rates (\code{ThetaSS}) can take positive values between 0 and 1. No subclass is assumed by the NPLCM models for SS data. We then use function \code{simulate_nplcm()} to produce a simulated data set:
\begin{verbatim}
R> J.BrS <- 6; J.SS <- 2; K <- 2
R> set_parameter_noreg <- list(Nd = 300,  Nu = 300, 
+     cause_list      = c("A","B","C","D","E","F"),
+     etiology        = c(0.5,0.2,0.15,0.05,0.05,0.05), 
+     meas_nm         = list(MBS = c("MBS1"),MSS=c("MSS1")), 
+     pathogen_BrS    = c("A","B","C","D","E","F"),
+     PsiBS           = cbind(c(0.25,0.25,0.2,0.15,0.15,0.15), 
+                             c(0.2, 0.2, 0.25,0.1,0.1,0.1)),
+     ThetaBS         = cbind(c(0.95,0.9,0.9,0.9,0.9,0.9), 
+                             c(0.95,0.9,0.9,0.9,0.9,0.9)),
+     Eta             = t(replicate(J.BrS,c(0,1))), 
+     Lambda          = c(0.5,0.5) ,           
+     pathogen_SS     = c("A","B"),
+     PsiSS           = c(0,0,NA,NA,NA,NA), 
+     ThetaSS         = c(0.15,0.1,NA,NA,NA,NA) 
+ )
R> data_nplcm_noreg   <- simulate_nplcm(set_parameter_noreg)$data_nplcm
\end{verbatim}

The data set can also be directly accessed by \code{data(data_nplcm_noreg)}. We can use the \code{summarize_BrS()} function to get summary statistics of the BrS measurements of the simulated data set. This function outputs the number of cases and controls, the observed marginal means for each measured item in the cases and the controls, along with the names of the measurements. The following line of code computes summaries for the single slice of the BrS data (\code{data_nplcm_noreg\$Mobs\$MBS[[1]]}) given the vector of case-control statuses (\code{data_nplcm_noreg\$Y}); another function \code{summarize_SS()} can be used similarly for SS measurements:
\begin{verbatim}
R> summarize_BrS(data_nplcm_noreg$Mobs$MBS[[1]],data_nplcm_noreg$Y)
R> summarize_SS(data_nplcm_noreg$Mobs$MSS[[1]],data_nplcm_noreg$Y)
\end{verbatim}

 In addition to producing quick summary statistics, \pkg{baker} also provides functionalities to organize and store  pertinent information about the BrS (or SS) measurements. For example, in the context of the PERCH study, the specimen name can be saved in the \code{specimen} argument, e.g., \code{"NP"} for a nasopharyngeal specimen. Another piece of information can be saved in the \code{test} argument, e.g. \code{"PCR"} (polymerase chain reaction) - here we use \code{"1"} for illustration; \code{quality} specifies the measurement quality (e.g. \code{"BrS"} or \code{"SS"}). The output of this function can be found in Appendix \ref{output::brs_object}. 
\begin{verbatim}
R> BrS_object_1 <- make_meas_object(patho = set_parameter_noreg$pathogen_BrS, 
+    specimen = "MBS", test = "1", quality = "BrS", cause_list =
+    set_parameter_noreg$cause_list)
R> SS_object_1 <- make_meas_object(patho=LETTERS[1:J.SS],
+   "MSS","1","SS",set_parameter_noreg$cause_list)
R> clean_options <- list(BrS_objects = make_list(BrS_object_1), 
+   SS_objects  = make_list(SS_object_1))
\end{verbatim}

\subsection{Simulate data with covariates}
\label{sec:simulate_with_covariates}
In Appendix \ref{sec:appendix_simulate_X}, we provide 1) example code to simulate data \code{data_nplcm_reg_nest_strat} with two subclasses (``nested'') and two covariate strata; and 2) code to load a pre-simulated data set with a continuous covariate and a two-level discrete covariate: \code{data(data_nplcm_reg_nest)}. We refer the reader \code{simulate_nplcm()} in the help files of \pkg{baker} for more examples of how to simulate data with individual-level covariates.

\subsection{Specifying models}
\label{sec:model_specification}
We provide examples of \code{model_options} by specifying four models:
\begin{itemize}
    \item \code{model_options_no_reg};
    \item \code{model_options_no_reg_with_SS};
    \item \code{model_options_reg_nest_strat};
    \item \code{model_options_reg_nest},
\end{itemize}
the first two of which does not perform regression and the last two perform regression. Various NPLCMs can be specified via three named lists:
\begin{itemize}
    \item \code{use_measurements}:
    can be \code{"BrS"} or \code{"SS"} or \code{c("BrS","SS")} to represent the quality of the data sources used for model fitting;
    \item \code{likelihood}: a named list defining the model likelihood of the desired NPLCM;
    \item \code{prior}: a named list defining the prior distributions for the associated parameters.
\end{itemize} 

\subsubsection{Specify models without regression}

In the code example below, \code{use_measurements = "BrS"} indicates that only the bronze-standard data are used in fitting an NPLCM. For specifying the likelihood, \code{k_subclass} indicates whether or not to use the conditional independence model (\code{k_subclass = 1} corresponds to the model with conditional independence given a cause, referred to as ``non-nested model"). \code{Eti_formula} specifies the regression formula for relating the CSCFs to case covariates; \code{FPR_formula} specifies the regression formula for relating the subclass weights to covariates (must be common to case and control subjects). In terms of the prior distribution (\code{prior}), \code{Eti_prior}
here specifies a numeric vector of length equal to the number of causes; this vector are Dirichlet hyperparameter for the population CSCFs. The \code{TPR_prior} specifies informative priors for the true positive rates; for example, we can specify a range \code{"0.55"} to \code{"0.99"} for \code{MBS1} measurement. 
\begin{verbatim}
R> cause_list <- c("A","B","C","D","E","F")
R> model_options_no_reg <- list(use_measurements = c("BrS"), 
+  likelihood   = list(cause_list = cause_list, k_subclass = 2,
+    Eti_formula = ~-1, FPR_formula = list(MBS1 = ~-1)),
+  prior= list(Eti_prior = overall_uniform(1,cause_list),
+    TPR_prior  = list(BrS = list(info  = "informative", input = "match_range",
+     val = list(MBS1 = list(up =  list(rep(0.99,J.BrS)), 
+                           low = list(rep(0.55,J.BrS)))) ))))     
\end{verbatim}

The main function (\code{nplcm()}) will use another built-in function \code{assign_model()} to check the \code{model_options} argument against the \code{data_nplcm} argument and will return information about the desired NPLCM. Users can use this to check that they have set up their \code{model_options} correctly. In particular, the output from the following code snippet can be found in Appendix \ref{output::assign_model_output}. 
\begin{verbatim} 
R> assign_model(model_options_no_reg,data_nplcm_noreg)
\end{verbatim}

In addition, we can specify a model that uses both BrS and SS data to fit an NPLCM without regression. To do this, we just need to modify the \code{data_nplcm} argument by adding SS data to the list \code{data_nplcm\$Mobs\$MSS} and specify a TPR prior for the SS data, e.g., \code{"0.01"} to \code{"0.5"} for \code{"MSS1"} measurements. This flexibility shows the package can work with multiple sources of data. 
\begin{verbatim}
R> model_options_no_reg_with_SS <- model_options_no_reg
R> model_options_no_reg_with_SS$use_measurements <- c("BrS","SS")
R> model_options_no_reg_with_SS$prior$TPR_prior$SS <-
  list(info  = "informative", input = "match_range",
   val   = list(MSS1 = list(up = list(rep(0.5,length(SS_object_1$patho))),
                           low = list(rep(0.01,length(SS_object_1$patho))))))
\end{verbatim}

\subsubsection{Specify models for regression analyses}
To set the \code{model_options} argument with regression covariates, we need to modify the \code{Eti_formula} and \code{FPR_formula} arguments. Here we use simulated data for illustration (see the final line of Appendix \ref{sec:appendix_simulate_X}). Recall that we do not let TPR vary by covariates in standard NPLCMs. In the following, because all the covariates are discrete, we specify symmetric Dirichlet priors with hyperparameters \code{1}s for the vector of CSCFs in each stratum of the covariate \code{SITE}. Other priors that have been set to defaults include the smoothness selection hyperparamters for the case and control FPRs (usually taken to be non-informative, given that FPRs can be estimated from the data).
\begin{verbatim}
R> model_options_reg_nest_strat <-  list(use_measurements = c("BrS"), 
+  likelihood  = list(cause_list = cause_list,
+   k_subclass = 2, Eti_formula = ~ -1+as.factor(SITE),
+   FPR_formula = list(MBS1 =  ~ -1 + as.factor(SITE))),
+  prior= list(Eti_prior  = c(2,2),
+   TPR_prior  = list(BrS = list(info  = "informative", input = "match_range", 
+     val = list(MBS1 = list(up =  list(rep(0.99,J.BrS)),
+                           low = list(rep(0.55,J.BrS)))) ))))
\end{verbatim}

Again, we can check that we have set up the NPLCM correctly using \code{assign_model()}. 

To fit the same model as above, but with an additional continuous covariate \code{"DATE“}, all we need to do is modify the regression formula:

\begin{verbatim}
R> model_options_reg_nest <- model_options_reg_nest_strat
R> model_options_reg_nest$likelihood$Eti_formula <- 
+  ~ -1+s_date_Eti(DATE,Y,basis='ps',dof=7)+as.factor(SITE)
R> model_options_reg_nest$likelihood$FPR_formula <- 
+   list(MBS1 =  ~ -1 +s_date_FPR(DATE,Y,basis = "ps",dof=5) + as.factor(SITE))
\end{verbatim}

\subsection{Setting up MCMC}
\label{sec:mcmc_options}

Finally, we specify the \code{mcmc_options} argument of \code{nplcm()}, including the number of chains (\code{n.chains}), the number of total iterations (\code{n.itermcmc}), the number of burn-in iterations (\code{n.burnin}), the  thinning interval (\code{n.thin}), whether or not individual level latent class predictions are desired (\code{individual.pred = TRUE} or \code{FALSE}) and whether or not to sample from the posterior predictive distributions (\code{ppd = TRUE} or \code{FALSE}). In addition, we must specify the path to the directory where the model should write the posterior samples (\code{result.folder}) and the path to the directory of the \code{.bug} model files (\code{bugsmodel.dir}). 
\begin{verbatim}
R> thedir <- paste0(tempdir(),"_no_reg"); dir.create(thedir)
R> dput(data_nplcm_noreg,file.path(thedir,"data_nplcm.txt"))   
R> dput(clean_options, file.path(thedir,"data_clean_options.txt"))
R> mcmc_options_no_reg <- list( n.chains = 3, n.itermcmc = 2000,  
+   n.burnin = 1000, n.thin = 1, individual.pred = TRUE, 
+   ppd = TRUE, result.folder = thedir, bugsmodel.dir = thedir)
\end{verbatim}

Now that we have specified all the required arguments for \code{nplcm()}, we can fit our NPLCM using the following code: 

\begin{verbatim}
R> nplcm_noreg <- nplcm(data_nplcm_noreg, model_options_no_reg,
+    mcmc_options_no_reg)
\end{verbatim}

We can similarly obtain the other fitted models: \code{nplcm\_noreg\_with\_SS}, \code{nplcm\_reg\_nest\_strat}, and \code{nplcm_reg_nest}; see Appendix \ref{sec:appendix_model_fits} for three separate uses of the main function \code{nplcm()}.

Because we use \proglang{JAGS} for automatic and versatile Bayesian inference of the models, we need to supply \proglang{JAGS} with \code{.bug} files with the requested forms of model likelihood and prior distribution. In the above code, \code{nplcm()} automatically interprets the specified model options and writes them in \code{.bug} files. In this case, \code{nplcm()} uses some internal function to generate a \code{.bug} model file for conditional independence models without regression, using BrS data. This model file will be stored in the path specified by \code{result.folder}. In addition, \code{fs::dir\_tree(path = nplcm\_noreg\$DIR\_NPLCM, recurse = TRUE)} lists all the resulting files in the folder storing model results. The output of \code{nplcm()} is an \proglang{S3} object of class \code{nplcm} which may be used by generic methods such as \code{summary()}, \code{print()}, and \code{plot()} (see Section \ref{sec:generic}).

\subsection{Convergence and model diagnostics}
\label{sec:postprocess}

After MCMC iterations are completed, we can assess convergence of the sampling chains for parameters using recorded information. The file \code{jagsdata.txt} contains all the data and pre-specified hyperparameters used when fitting the model. The files \code{CODAchain1.txt} and \code{CODAindex.txt} together record the posterior samples. Here we illustrate by using \code{pEti}, which stores the posterior samples of the CSCFs; we illustrate by focusing on model results obtained from an NPLCM without regression \code{nplcm_noreg}. Our package does not provide built-in functions to assess convergence and mixing of the sampling algorithm; in the following, we illustrate how to extract sampled values based on the outputs of \code{nplcm()}. However, \pkg{baker} provides built-in functions for performing posterior predictive checking. Posterior samples can be read into the \code{coda} format using the \pkg{coda} package. In particular, to retrieve the posterior samples, recall that the output from the \code{nplcm()} function contains the path to the directory where we stored the posterior samples and can be called using \code{\$DIR_NPLCM}, as shown in the following example: 
\begin{verbatim}
R> res_nplcm_noreg <- coda::read.coda(file.path(nplcm_noreg$DIR_NPLCM,
"CODAchain1.txt"), file.path(nplcm_noreg$DIR_NPLCM,"CODAindex.txt"),quiet=TRUE)
R> get_res   <- function(x,res) res[,grep(x,colnames(res))]
R> res <- get_res("pEti",res_nplcm_noreg)
\end{verbatim}

\subsubsection{Convergence diagnostics}

The posterior samples obtained above can be further manipulated using external packages that provide algorithm convergence diagnostic functionalities. For example, \newline \code{coda::raftery.diag(mcmc.list(res))} produces the Raftery diagonistic for convergence and the effective sample sizes for our model parameters. As another example,\newline \code{ggmcmc::ggs_traceplot(ggmcmc::ggs(res))} plots the sampling trajectories for parameters of interest. See \citet{brooks2011handbook} for a more complete review of approaches and considerations for convergence assessment.






\subsubsection{Model diagnostics}

The \pkg{baker} package provides two useful posterior predictive checking functions. For each slice of BrS measurement data:

\begin{itemize}
    \item Standardized log odds ratio differences (SLORD); this is based on pairwise associations among the BrS measurements: near-zero SLORDs indicate the association is adequately characterized by the model;
    \item Probabilities of multivariate binary patterns with the highest frequencies; this is based on all the dimensions of measurements, not just pairwise, hence providing additional capability to assess the adequacy of the model in capturing the observed frequencies of multiple binary patterns.
\end{itemize}

Because the posterior predictive samples are stored in \code{\$DIR_NPLCM} (when \code{ppd = TRUE} was set in the MCMC options), one can read in these samples and perform desired posterior predictive checking with other statistics if desired.

First, the following code snippet outputs a figure of posterior predicted pairwise LOR (log-odds ratios) compared with the observed LOR for the BrS data. 
\begin{verbatim}
R> plot_check_pairwise_SLORD(nplcm_noreg$DIR_NPLCM,slice=1)
\end{verbatim}

Second, the following code snippet produces a figure shown in Figure \ref{fig:plot_check_common_patterns} of Appendix  based on the \proglang{S3} object of class \code{nplcm} (\code{"nplcm_noreg"}), for the first slice of the bronze-standard data (\code{slice_vec = 1}), and \code{npat = 5} patterns with the highest frequency in the observed data:
\begin{verbatim}
R> plot_check_common_pattern(list(nplcm_noreg$DIR_NPLCM),slice_vec=1,n_pat=5)
\end{verbatim}

\subsection{Summary, plot, print}
\label{sec:generic}

The \pkg{baker} package provides \code{summary()} methods to produce quick numerical summaries of the model specifications, posterior means and $95\%$ credible intervals for population CSCFs. For example, \code{summary(nplcm_no_reg)} shows information about the result obtained from an NPLCM without regression; the actual outputs are shown in the Appendix \ref{sec::appendix_summary_withSS}. 

The generic \code{plot()} methods in \pkg{baker} produces graphical summaries of the fitted model according to the type of NPLCMs fitted. For an NPLCM without regression, this will produce a multi-paneled figure that summarizes the data, prior and posterior. In our experience, the figure facilitates answering questions like ``where does the information come from?'', e.g., how the BrS and SS data summaries are consistent with the obtained posterior inferential results for each cause. Figure \ref{fig:base_plot_noreg} displays the output of the following code snippet. The detailed explanation of the details in the resulting figures can be found in the user manual of the \pkg{baker} package:
\begin{verbatim}
R> plot(nplcm_noreg_with_SS, bg_color = NULL)
\end{verbatim}

\begin{figure}[htp]
    \centering
    \includegraphics[width=\linewidth]{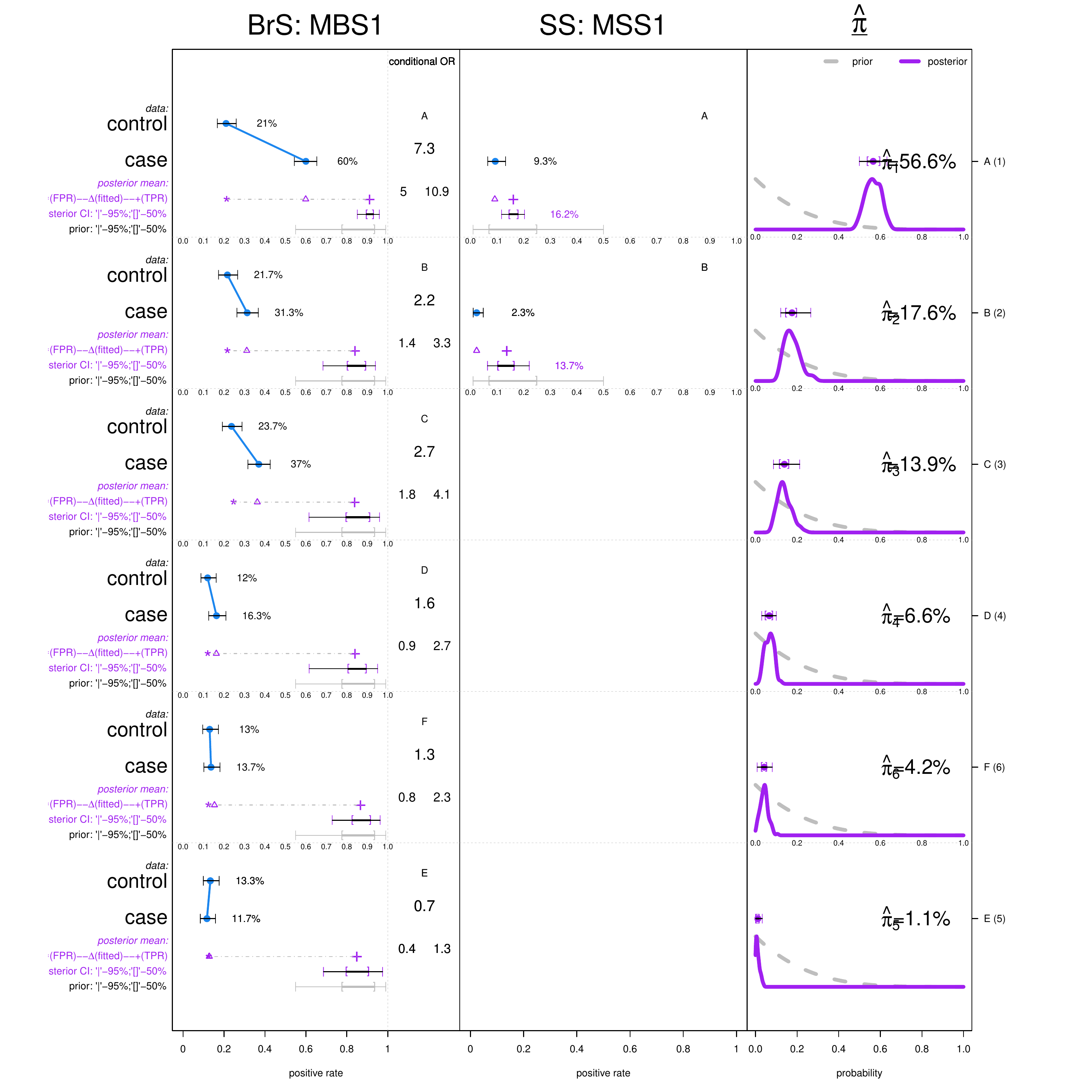}
    \caption{Output from \code{plot} for an \code{nplcm} object \code{"nplcm\_no\_reg\_with\_SS"}. The data, prior, and posterior summaries are displayed for each latent class in the rows; see \citet[Figure 3,]{wu2016partially} for additional figure descriptions.} 
    \label{fig:base_plot_noreg}
\end{figure}

\begin{figure}
    \centering
    \includegraphics[width=\linewidth]{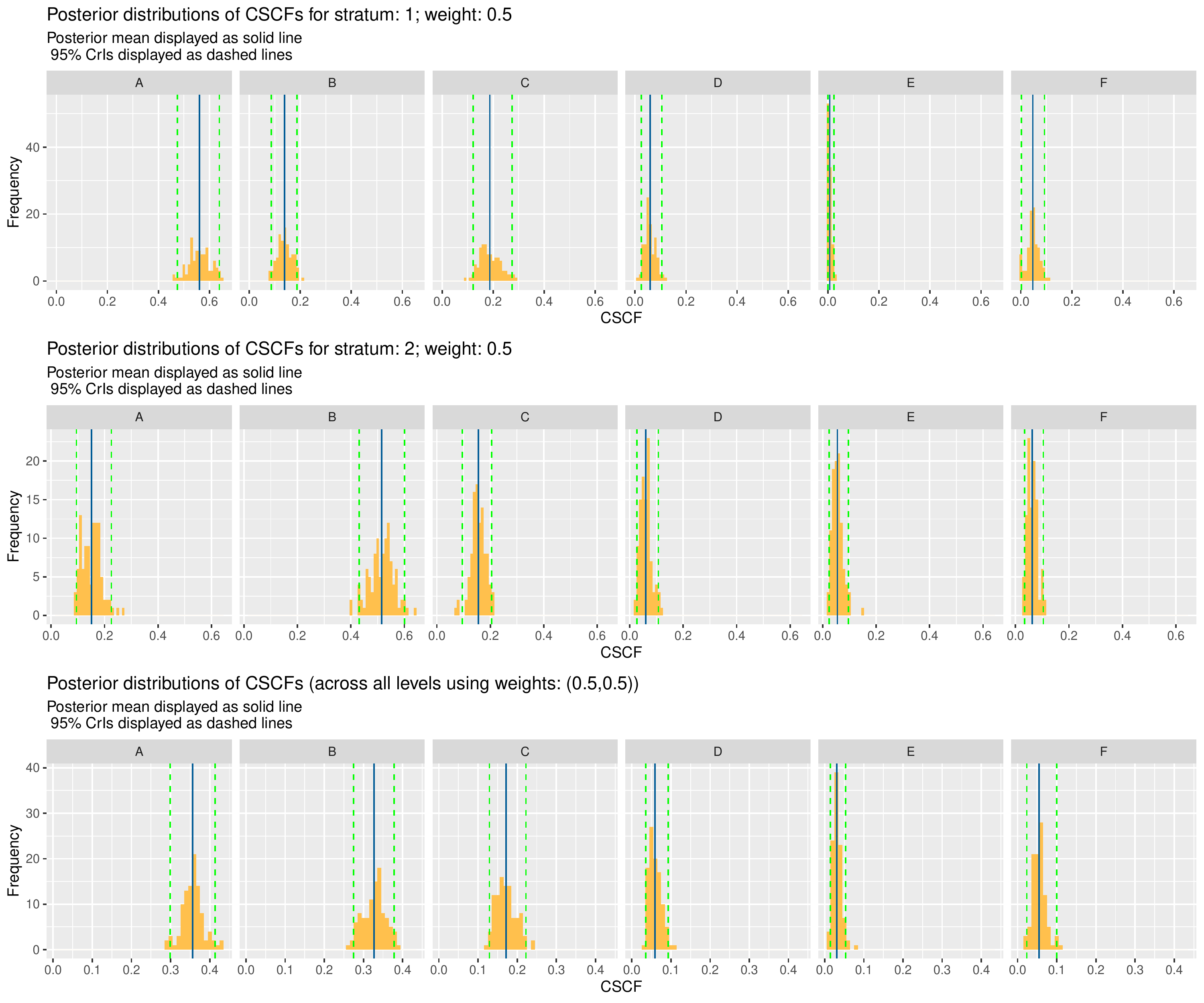}
    \caption{A graphical summary from \code{plot} of a fitted \code{nplcm} object \code{"nplcm\_reg\_nest\_strat"} with a two-level discrete covariate. The final row shows the marginal posterior distributions for causes \code{"A"} to \code{"F"} where stratum weights are 0.5 and 0.5; the weights default to empirical weights and can be user-specified. The other rows show the marginal posterior distributions of the CSCFs for each stratum.} 
    \label{fig:base_plot_discrete_reg}
\end{figure}

In the case of an NPLCM with one or more discrete covariates for CSCF regression (without continuous covariates), \code{plot()} will visualize the posterior distributions of the CSCFs for each covariate stratum. Figure \ref{fig:base_plot_discrete_reg} displays the output of the following code snippet: 
\begin{verbatim}
R> plot(nplcm_reg_nest_strat,show_levels = c(0,1,2))
\end{verbatim} 
 The above code produces a figure for each stratum (the final row is for the overall CSCF estimates as a weighted average across strata). To only plot the marginalized posterior distributions, we can set \code{show_levels = 0}.

In the case of an NPLCM with a continuous covariate, the \code{plot()} will produce a figure with two rows: one of the estimated marginal positive rates for cases and controls and one of the CSCFs for each disease class among the cases. We show an example of this plot in Figure \ref{fig:plot_continuous_reg} where CSCFs may vary by enrollment date (x-axis). If we have discrete covariates as well as a continuous covariate in the CSCF regression, we can make these plots for each stratum of the discrete covariates by including the stratum as an argument in the \code{plot} function:   
\begin{verbatim}
R> DISCRETE_BOOL <- data_nplcm_reg_nest$X$SITE == 1
R> plot(nplcm_reg_nest, stratum_bool = DISCRETE_BOOL)
\end{verbatim}


\begin{figure}
    \centering
    \includegraphics[width=\linewidth]{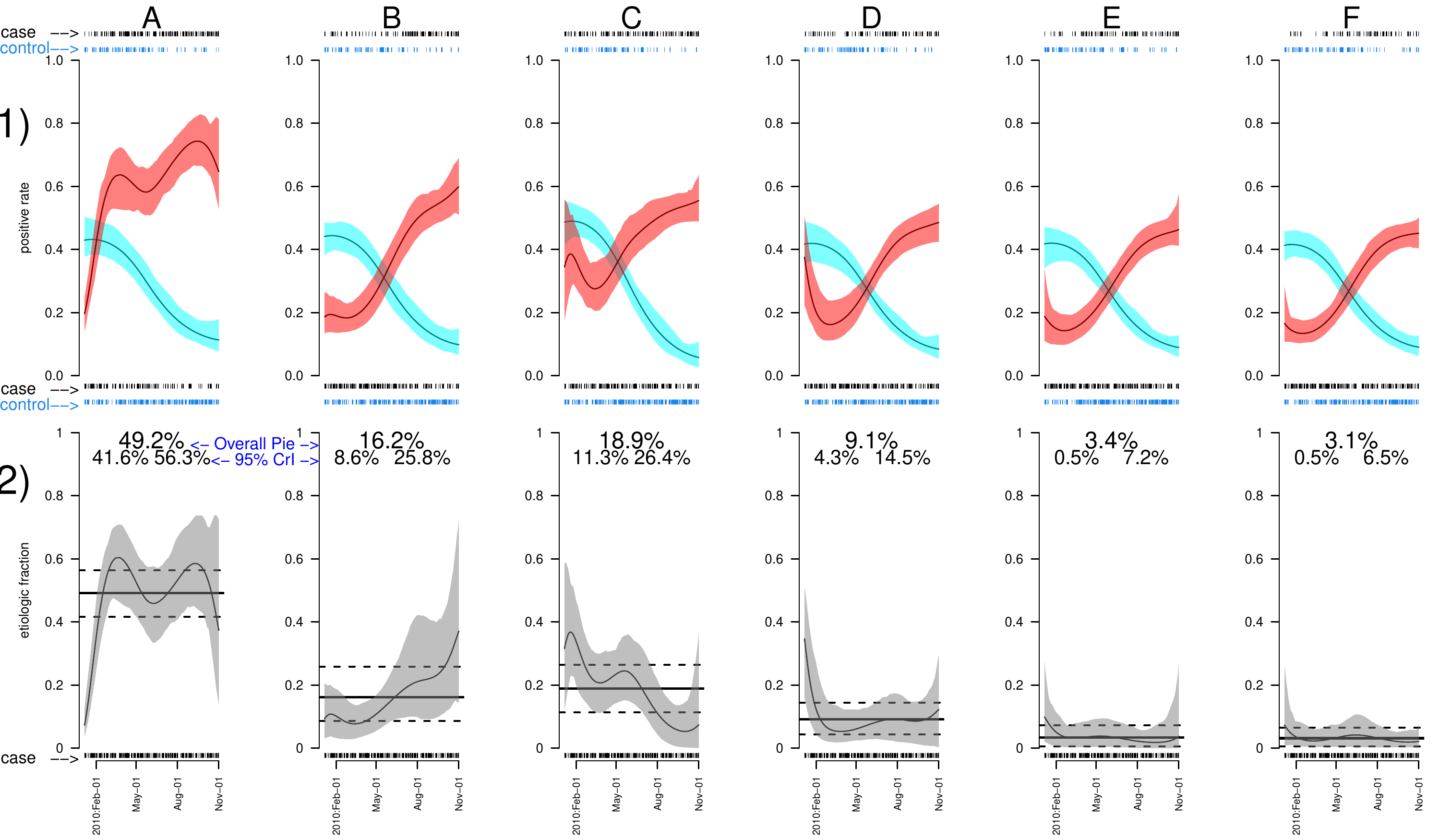}
    \caption{Output from \code{plot} for an \code{nplcm} object \code{"nplcm\_reg\_nest"} with a continuous regression covariate (seasonality). The top plot displays the estimated TPR for each latent class and the bottom plot displays the estimated CSCFs for each latent class. See for detailed description of the figure in \citet{wu2021regression}.} 
    \label{fig:plot_continuous_reg}
\end{figure}

\subsection{PERCH study example}
\label{sec:real_data_example}

Pneumonia is a clinical condition associated with infection of the lung tissue, which can be caused by multiple species of pathogens. In studies of pneumonia etiology, a cause is the subset of one or more pathogens infecting the lung. Knowledge about population-level cause-specific etiologic contributions can help prioritize prevention programs and design treatment algorithms. The Pneumonia Etiology Research for Child Health (PERCH) study is a seven-country case-control study of the etiology of severe and very severe pneumonia. \citep{o2019aetiology} The primary aim of the study is to estimate the etiologic contributions quantified by \textit{cause-specific case fractions} (CSCFs), which may vary by individual-level factors such as age, disease severity, nutrition status and human immunodeficiency virus (HIV) status. 

In the PERCH study, tabulating case frequencies by cause is infeasible, because the lung-infecting pathogen(s) can rarely be directly observed due to potential clinical complications associated with invasive lung aspiration procedure \citep{o2019aetiology}. As an alternative, a non-invasive real-time polymerase chain reaction (PCR) test was made on each case's nasopharyngeal (NP) specimen, outputting presence or absence of a list of pathogens in the nasal cavity. The NP multivariate binary measurements are imprecise indicators for what pathogens infected the lung. In particular, detecting a pathogen in a case's nasal cavity does not indicate it caused lung infection. To provide statistical control for false positive detections, the PERCH study also performed NPPCR tests on pneumonia-free controls.

We illustrate with a regression analysis with $518$ cases and $964$  controls from one of the PERCH study sites in the Southern Hemisphere that collected more complete information on age (dichotomized to younger or older than one year), HIV status (positive or negative), disease severity for cases (severe or very severe), and presence or absence of seven species of pathogens (five viruses and two bacteria, representing a subset of pathogens evaluated) in NPPCR. The names of the pathogens and the abbreviations are (i) bacteria:  \textit{Haemophilus influenzae} ({\sf HINF}) and \textit{Streptococcus pneumoniae} ({\sf PNEU}), (ii) viruses: adenovirus ({\sf ADENO}), human metapneumovirus type A or B ({\sf HMPV\_A\_B}), parainfluenza type 1 virus ({\sf PARA\_1}), rhinovirus ({\sf RHINO}), and respiratory syncytial virus ({\sf RSV}).   We also include in the analysis the case-only, perfectly specific but imperfectly sensitive blood culture (BCX) diagnostic test results for two bacteria from cases only. For BCX data, we assume perfect specificity which is guided by the fact that if a pathogen did not infect the lung, it cannot be cultured from the blood (so we do not need control data to estimate the specificities). Detailed analyses of the entire data are reported elsewhere \citep{perchresults}.

Since the PERCH study data is not yet public and freely accessible, we provide example code of running \code{nplcm()} with model specifications (\code{perch_model}) and model fitting options (\code{perch_mcmc}). To more fully illustrate the functionality of \pkg{baker}, we use a pre-run posterior analysis of PERCH data and demonstrate numerical and graphical summaries of a simple regression analysis using \pkg{baker}. Sampling for this model in \proglang{JAGS} took approximately 5 minutes on a machine with an Quad-Core Intel i7 2.9 GHZ CPU and 16 GB of RAM running OSX Version 12.3 Beta. 

First, we show the structure of the input data:
\begin{verbatim}
R> str(perch_data)
# List of 3
# $ Mobs:List of 2
# ..$ MBS:List of 1
# .. ..$ NPPCR:'data.frame':	1488 obs. of  7 variables:
#   .. .. ..$ HINF    : int [1:1488] 0 1 0 0 1 1 0 0 0 0 ...
# .. .. ..$ PNEU    : int [1:1488] 0 0 0 0 0 0 0 0 0 0 ...
# .. .. ..$ ADENO   : int [1:1488] 0 0 0 0 NA NA NA NA 0 1 ...
# .. .. ..$ HMPV_A_B: int [1:1488] 0 0 0 0 NA NA NA NA 0 0 ...
# .. .. ..$ PARA_1  : int [1:1488] 0 0 0 0 NA NA NA NA 0 0 ...
# .. .. ..$ RHINO   : int [1:1488] 0 0 0 0 NA NA NA NA 0 0 ...
# .. .. ..$ RSV     : int [1:1488] 0 0 0 0 NA NA NA NA 0 0 ...
# ..$ MSS:List of 1
# .. ..$ BCX:'data.frame':	1488 obs. of  2 variables:
#   .. .. ..$ HINF: int [1:1488] 0 0 0 0 0 0 0 0 0 0 ...
# .. .. ..$ PNEU: int [1:1488] 0 0 0 0 0 0 0 0 0 0 ...
# $ X   :'data.frame':	1488 obs. of  4 variables:
# ..$ patid          : chr [1:1488] "S00021" "S00023" "S00026" "S00027" ...
# ..$ AGE            : int [1:1488] 0 1 0 0 0 0 0 0 1 0 ...
# ..$ HIV2           : int [1:1488] 0 0 0 0 0 0 0 0 1 0 ...
# ..$ ALL_VS         : int [1:1488] 0 0 1 0 1 0 1 0 0 0 ...
# $ Y   : num [1:1488] 1 1 1 1 1 1 1 1 1 1 ...
\end{verbatim}

We specify \code{L = 8} causes comprised of seven singleton-pathogen causes along with a cause named \code{"other"} that represents a generic non-specified (``NoS'') cause. For BrS data, we use nasopharyngeal polymerase chain reaction (NPPCR), which results in case-control measurements upon \code{J.BrS} targeted pathogens (bacteria and viruses with abbreviated names \code{"HINF"}, \code{"PNEU"}, \code{"ADENO"}, \code{"HMPV_A_B"}, \code{"PARA_1"}, \code{"RHINO"}, \code{"RSV"}). For SS data, we use blood culture (BCX) results for two species of bacteria (\code{"HINF","PNEU"}). We then organize these measurement information into \code{BrS_object_1} and \code{SS_object_1}, respectively. Finally, we create a temporary folder for storing model results.

\begin{verbatim}
R> cause_list <- c("HINF","PNEU","ADENO","HMPV_A_B","PARA_1","RHINO","RSV","other")
R> patho_BrS_NPPCR <- c("HINF","PNEU","ADENO","HMPV_A_B","PARA_1","RHINO","RSV")
R> patho_SS_BCX <- c("HINF","PNEU")
R> BrS_object_1 <- make_meas_object(patho_BrS_NPPCR,"NP","PCR","BrS",cause_list)
R> SS_object_1  <- make_meas_object(patho_SS_BCX,"B","CX","SS",cause_list)
R> perch_clean <- list(BrS_objects = list(BrS_object_1),
+                       SS_objects = list(SS_object_1))
R> result_folder  <- tempdir()
R> dir.create(result_folder)
\end{verbatim}

We the specify an NPLCM with five subclasses. In addition, we 1) let population etiology (``CSCFs'') depend on age, severity status, and HIV status, and 2) let subclass weights depend on age and HIV status. We use both BrS and SS data for estimation. The TPR priors are specified via prior 2.5\% and 97.5\% prior Beta quantiles (0.5 to 0.9 for BrS measurement TPRs; 0.05 to 0.2 for SS measurement TPRs); these prior ranges were elicited from domain scientists. This can be achieved by the following code.
\begin{verbatim}
R> perch_model <- list(use_measurements = c("BrS","SS"),  
+ likelihood  = list(cause_list = cause_list, k_subclass = c(5),                    
+   Eti_formula = ~ -1+as.factor(AGE)+as.factor(ALL_VS)+as.factor(HIV2),             
+   FPR_formula = list(NPPCR =  ~ -1+as.factor(AGE)+as.factor(HIV2) )),  
+ prior = list(Eti_prior  = c(2,2),  
+   TPR_prior   = list(
+    BrS = list(info  = "informative", input = "match_range",
+      val = list(NPPCR = list(up = list(rep(0.9,length(BrS_object_1$patho))),
+                        low = list(rep(0.5,length(BrS_object_1$patho)))  ))),
+    SS = list(info = "informative", input = "match_range",
+      val = list(MSS1 = list(up = list(rep(0.2,length(SS_object_1$patho))),
+                  low = list(rep(0.05,length(SS_object_1$patho))) ))))))
\end{verbatim}

We then specify the settings for the MCMC algorithm (\code{perch_mcmc}):
\begin{verbatim}
R> perch_mcmc <- list(n.chains   = 1, n.itermcmc = 200, n.burnin   = 100,
+    n.thin = 1, individual.pred = TRUE, ppd = TRUE,
+    result.folder = result_folder, bugsmodel.dir = result_folder)
\end{verbatim}

Below, we fit the specified model. Here we assume the data set  (\code{data_nplcm}) has been loaded from the real data (not yet publicly available). We store data (to \code{"data_nplcm.txt"}), store data cleaning information (to \code{"data_clean_options.txt"}) check model specifications via \code{assign_model()}, and fit the model by \code{nplcm()}:
\begin{verbatim}
R> dput(perch_data,file.path(perch_data$result.folder,"data_nplcm.txt"))
R> dput(perch_clean,file.path(perch_mcmc$result.folder,"data_clean_options.txt"))
R> assign_model(perch_model,perch_data)
R> rjags::load.module("glm")
R> perch_fit <- nplcm(data_nplcm,model_options,perch_mcmc)
\end{verbatim}

We can summarize the fitted object using \code{summary(perch_fit)} (results not shown here). We can also visualize rich information about the posterior distribution of the population-level CSCFs in the cases using generic function \code{plot()} which shows the high population level etiologic importance of the virus {\sf RSV}:
\begin{verbatim}
R> plot(perch_fit)
\end{verbatim}

\begin{figure}[htp]
    \centering
    \includegraphics[width=\linewidth]{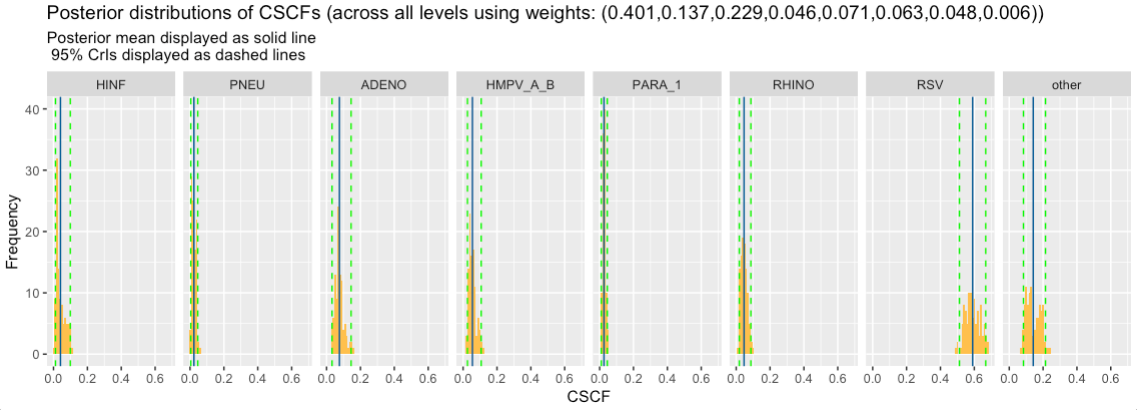}
    \caption{Marginal posterior distributions for each of \code{L = 8} pre-specified causes as visualized by the output from \code{plot} for an \code{nplcm} object \code{"perch\_fit"}.} 
    \label{fig:perch_plot}
\end{figure}

As mentioned before, \pkg{baker} can produce versatile posterior inferences about various unknown quantities based on the posterior samples. In particular, we can estimate the posterior probabilities of class membership probabilities for each individual case given each individual case’s observed measurements:
\begin{verbatim}
R> get_individual_prediction(perch_fit)
#       HINF PNEU ADENO HMPV_A_B PARA_1 RHINO  RSV other
#  [1,] 0.03 0.00  0.02     0.03   0.04  0.03 0.49  0.36
#  [2,] 0.59 0.01  0.06     0.04   0.02  0.05 0.11  0.12
#  [3,] 0.05 0.01  0.07     0.06   0.02  0.05 0.71  0.03
#  [4,] 0.04 0.00  0.02     0.02   0.01  0.01 0.59  0.31
#  [5,] 0.10 0.00  0.02     0.03   0.02  0.07 0.76  0.00
\end{verbatim}

Here we show the posterior probabilities of eight disease classes (columns) for five random cases (rows). For each individual, the probabilities differ across causes indicating varying posterior etiologic importance. For each cause, subjects differ in the BrS and SS measurements and covariates, leading to distinct posterior class membership probabilities.

\section{Summary}
\label{sec:conclusions}

The \pkg{baker} package provides functionalities to estimate a suite of NPLCM-based models \citep{wu2016partially, wu2017nested, wu2021regression} for multivariate binary responses that are observed under a case-control design. \pkg{baker} has three major strengths: 1) it enables case-control analyses with or without covariates in the NPLCM framework, 2) it relaxes the ``conditional independence'' assumption often used in latent class analyses, and 3) it is designed to handle multiple sets of case-control or case-only measurements of distinct quality. Model results, posterior uncertainty assessment, and model diagnostics can also be readily summarized by \pkg{baker}.



Our future aim is to accommodate mixed categorical, ordinal, and continuous responses. The latent class model is naturally suited for multivariate discrete responses. We can extend our framework to handle multiple response levels by adding additional response probability parameters for each response level. Second, to accommodate ordinal and continuous responses, e.g., those produced by measurement technologies such as antibody titers and real-time PCR, class-specific mixture component likelihood functions may be specified via latent random Gaussian vectors, a subset of which are then linked to non-continuous responses via thresholding. Fast computational techniques for sampling Gaussian covariance matrices in multivariate probit models akin to \citet{zhang2021large} can be implemented. Finally, our current functions for NPLCM regression analyses will be expanded to accommodate higher dimensional covariates for CSCF and subclass weights using sparse Bayesian Additive Regression Trees \citep{linero2018bayesian}.


\section*{Acknowledgments}

\vspace{-.25cm}
This work was supported by a Michigan Institute of Data Science (MIDAS) seed grant (to Z.W.); the Patient-Centered Outcomes Research Institute (PCORI) Award [ME-1408-20318 to Z.W. and S.L.Z.]; and the National Institutes of Health grants [P30CA046592 to Z.W.]. We thank the PERCH study team led by Katherine O'Brien for providing the data and scientific advice, Maria Deloria-Knoll and Christine Prosperi for valuable feedback about \pkg{baker} and Jing Chu for preliminary simulations. We also thank John Kubale for valuable feedback.
\bibliography{software_ref}

\appendix
\section{Additional code to simulate data with covariates}
\label{sec:appendix_simulate_X}
\begin{verbatim}
R> N.SITE     <- 2                         
R> N          <- N.SITE*(300+300)           
R> 
R> CSCF_allsites <- list(c(0.5,0.2,0.15,0.05,0.05,0.05),
R>                           c(0.2,0.5,0.15,0.05,0.05,0.05))
R> 
R> out_list <- lapply(1:N.SITE,function(siteID){ 
R>   set_parameter <- list(    
R>     cause_list   = c("A","B","C","D","E","F"),
R>     etiology     = CSCF_allsites[[siteID]],
R>     pathogen_BrS = LETTERS[1:J.BrS],
R>     SS           = TRUE,   
R>     pathogen_SS  = c("A","B"),
R>     meas_nm      = list(MBS = c("MBS1"),MSS=c("MSS1")),
R>     Lambda       = c(0.5,0.5),  # control subclass weight for BrS
R>     Eta          = t(replicate(J.BrS,c(0,1))),  
R>     PsiBS        = cbind(c(0.25,0.25,0.2,0.15,0.15,0.15), 
R>                          c(0.2, 0.2, 0.25,0.1,0.1,0.1)), 
R>     PsiSS        = cbind(rep(0,J.BrS),rep(0,J.BrS)), 
R>     ThetaBS      = cbind(c(0.95,0.9,0.9,0.9,0.9,0.9), 
R>                          c(0.95,0.9,0.9,0.9,0.9,0.9)),
R>     ThetaSS      = cbind(c(0.25,0.10,0.15,0.05,0.15,0.15),
R>                          c(0.25,0.10,0.15,0.05,0.15,0.15)),
R>     Nd = 300,
R>     Nu = 300
R>   )
R>   out     <- simulate_nplcm(set_parameter)
R>   res   <- out$data_nplcm
R>   res$X <- data.frame(SITE=rep(siteID,(set_parameter$Nd+set_parameter$Nu)))
R>   return(res)
R> })
R> data_nplcm_unordered  <- combine_data_nplcm(out_list)
R> data_nplcm_reg_nest_strat <- subset_data_nplcm_by_index(data_nplcm_unordered,
R>                                     order(-data_nplcm_unordered$Y))
R> # load another data in `baker` with a continuous covariate
R> data(data_nplcm_reg_nest)   
\end{verbatim}

\section{Additional code to fit models}
\label{sec:appendix_model_fits}
\begin{verbatim}
R> nplcm_noreg_with_SS <- nplcm(data_nplcm_noreg,
+   model_options_no_reg_with_SS,mcmc_options_no_reg_with_SS)
R> nplcm_reg_nest_strat <- nplcm(data_nplcm_reg_nest_strat,
+   model_options_reg_nest_strat,mcmc_options_reg_nest_strat)
R> nplcm_reg_nest <- nplcm(data_nplcm_reg_nest,
+   model_options_reg_nest,mcmc_options_reg_nest)
\end{verbatim}

\section{Selected code outputs}

\subsection{Organized BrS data meta-information}
\begin{verbatim}
R> BrS_object_1
# $quality
# [1] "BrS"
# $patho
# [1] "A" "B" "C" "D" "E" "F"
# $name_in_data
# [1] "A_MBS1" "B_MBS1" "C_MBS1" "D_MBS1" "E_MBS1" "F_MBS1"
# $template
#      [,1] [,2] [,3] [,4] [,5] [,6]
# [1,]    1    0    0    0    0    0
# [2,]    0    1    0    0    0    0
# [3,]    0    0    1    0    0    0
# [4,]    0    0    0    1    0    0
# [5,]    0    0    0    0    1    0
# [6,]    0    0    0    0    0    1
# [7,]    0    0    0    0    0    0
# $specimen
# [1] "MBS"
# $test
# [1] "1"
# $nm_spec_test
# [1] "MBS1"
\end{verbatim}
\label{output::brs_object}

\subsection{Specifying model}
\label{output::assign_model_output}
\begin{verbatim}
R> assign_model(model_options_no_reg,data_nplcm_noreg)
# $num_slice
# MBS MSS MGS 
#   1   0   0 
# $nested
# [1] TRUE
# $regression
# $regression$do_reg_Eti
# [1] FALSE
# $regression$do_reg_FPR
#  MBS1 
# FALSE 
# $regression$is_discrete_predictor
# $regression$is_discrete_predictor$Eti
# [1] FALSE
# $regression$is_discrete_predictor$FPR
#  MBS1 
# FALSE 
\end{verbatim}

\subsection{Summary}
\label{sec::appendix_summary_withSS}
\begin{verbatim}
R> summary(nplcm_noreg_with_SS)
# [baker] summary: model structure 
#            fitted type:  no_reg 
# ---
#      name measurements:  MBS MSS MGS 
# slices of measurements:  1 1 0 
#                 nested:  TRUE 
# ---
#             regression:  
#                   etiology:  FALSE 
#                   name FPR:  MBS1 
#                        FPR:  FALSE 
# ---
# all discrete predictor:  
#                   etiology:  FALSE 
#                   name FPR:  MBS1 
#                        FPR:  FALSE 
# 
# ------- posterior summary -----------
#    post.mean     post.sd      CrI_025   CrI_0975
# A 0.56573053 0.036129324 0.4992234000 0.63213780
# B 0.17558873 0.037365233 0.1216957500 0.26571990
# C 0.13901539 0.032986620 0.0864293800 0.21263615
# D 0.06632659 0.021147999 0.0300819800 0.10015262
# E 0.01137363 0.009374251 0.0003651512 0.03293634
# F 0.04196514 0.019310869 0.0084014205 0.08096354
\end{verbatim}

\subsection{Posterior predictive checking}


\begin{figure}[htp]
    \centering
    \includegraphics{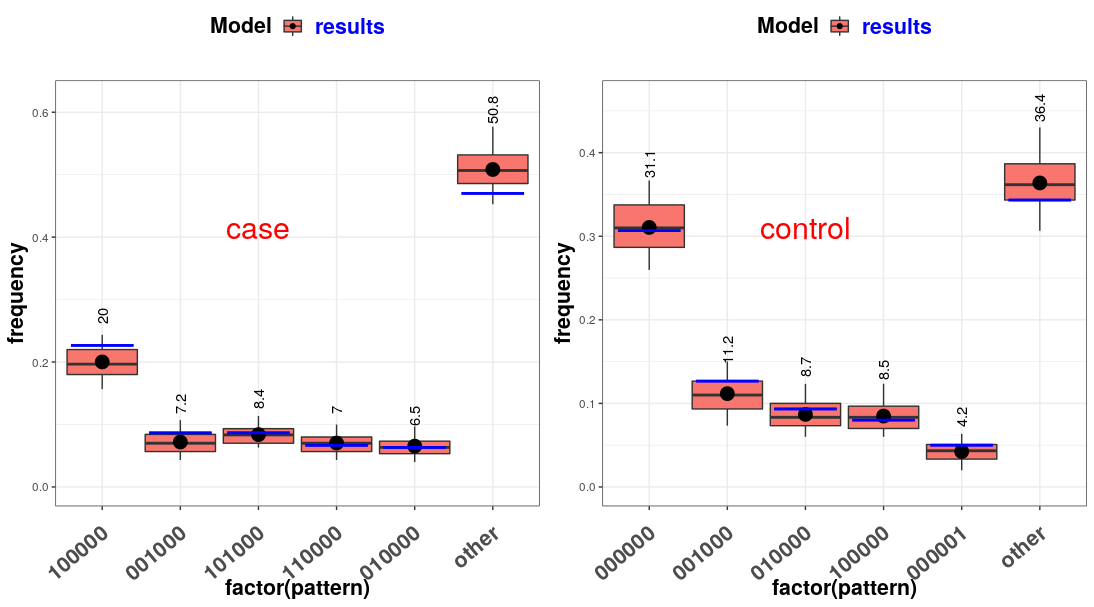}
    \caption{For the cases and the controls, posterior predictive checking based on the probability of multivariate binary patterns (five top patterns in the actual data and the rest aggregated). For each pattern, the posterior predictive distribution is shown by a boxplot; the horizontal blue bar  indicates the actual observed frequency. A large deviation of a horizontal bar from the corresponding boxplot suggests potential model misfit.}
    \label{fig:plot_check_common_patterns}
\end{figure}

\end{document}